\documentclass[%
 reprint,
 amsmath,amssymb,
 aps,
]{revtex4-1}
\usepackage{amsmath,times,amssymb,latexsym,graphics,braket,color,here}
\usepackage{graphicx}
\usepackage{dcolumn}
\usepackage{bm}
\usepackage{braket}
\usepackage{graphicx} 
\usepackage{color}   

\usepackage{hyperref}

\newcommand{\ex}[1]{\mathrm{e}^{#1}}

\newcommand{\pa}[1]{\left(#1 \right)}

\newcommand{\bb}[1]{\mathbb{#1}}

\newcommand{\ca}[1]{\mathcal{#1}}

\newcommand{\kett}[1]{\ket{#1\rangle}}

\newcommand{\fr}{\frac}

\def\be{\begin{equation}}
\def\ee{\end{equation}}
\def\ba{\begin{eqnarray}}
\def\ea{\end{eqnarray}}

 \def\a{{\alpha}}
 \def\ba{{\bar{\alpha}}}

\def\dd{{\mathrm{d}}}

\makeatletter
\let\cat@comma@active\@empty
\makeatother

\begin{document}
\preprint{RIKEN-iTHEMS-Report-22}
\title{Semiclassical Gravity from Averaged Boundaries in two-dimensional BCFTs}
\date{\today}
\author{Yuya Kusuki}\email[]{ykusuki@caltech.edu}
\affiliation{\it Walter Burke Institute for Theoretical Physics, 
California Institute of Technology, Pasadena, CA 91125, USA.}
\affiliation{\it RIKEN Interdisciplinary Theoretical and Mathematical Sciences (iTHEMS),
Wako, Saitama 351-0198, Japan.}

\begin{abstract}
The interpretation of semiclassical gravity as an ensemble-average of CFTs has provided much progress in understanding the factorization problem and the information paradox.
In this article, we consider an averaging over boundary conditions in two-dimensional boundary conformal field theories (BCFTs).
We show that the boundary averaging plays an important role not only in producing wormhole contributions but also in resolving the following problem.
In general setups in AdS/BCFT, one can find unphysical solutions where branes have (self-)intersections.
We show that the averaging nicely resolves this intersection problem.
This provides another piece of evidence that averaging plays an important role in reproducing Einstein gravity.
We also propose that the averaging provides a BCFT dual of the island model.
\end{abstract}
\maketitle

\section{Introduction}
\begin{figure}[t]
 \begin{center}
  \includegraphics[width=9.0cm,clip]{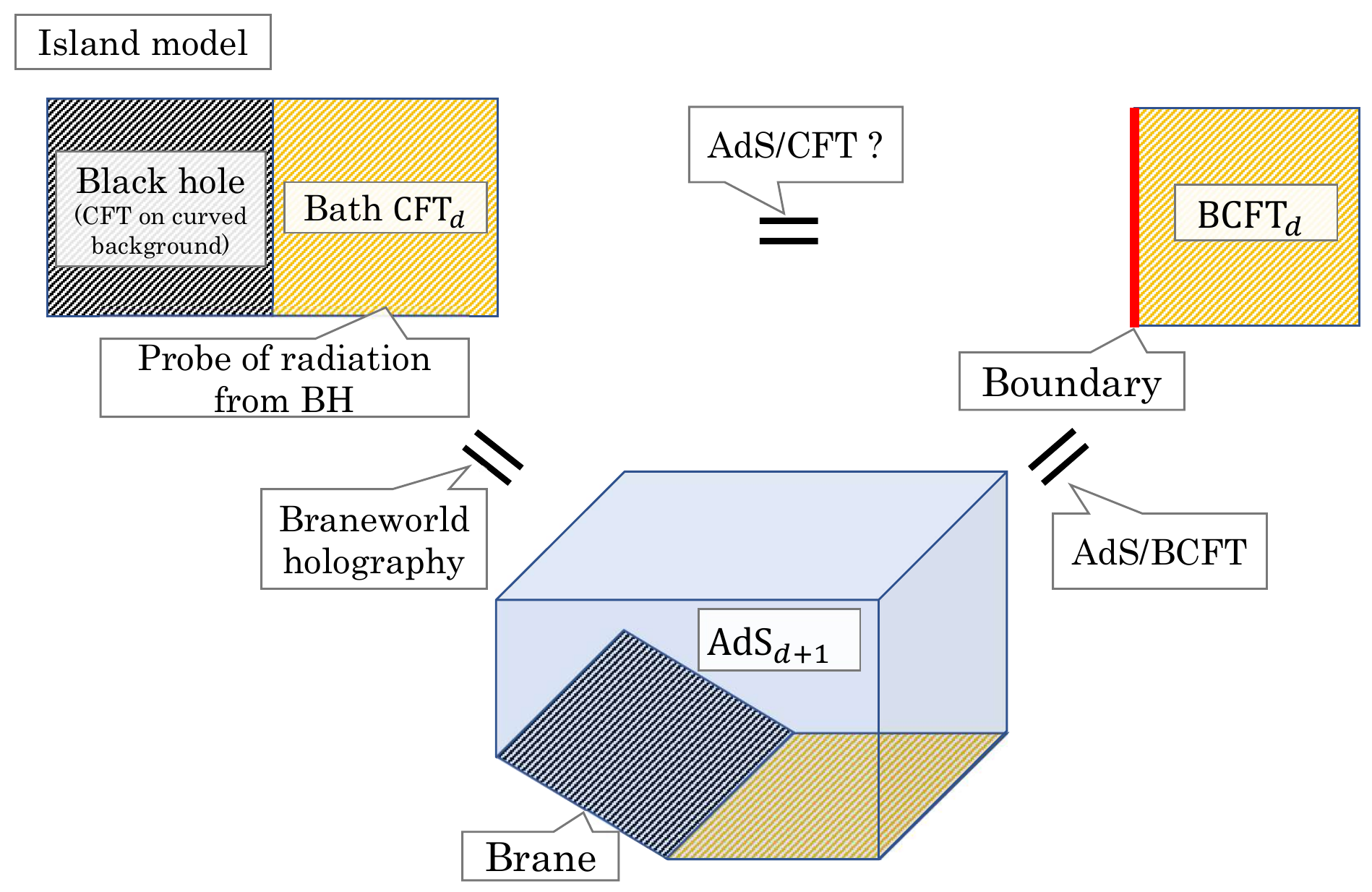}
 \end{center}
 \caption{The relation between (Left) the island model, (Center) AdS${}_{d+1}$ with end of the world (ETW) brane, and (Right) BCFT${}_d$.
 In AdS/BCFT, the intersection between the brane and the asymptotic boundary corresponds to the boundary of BCFT.
 From the viewpoint of JT gravity/averaged SYK correspondence, the boundary of BCFT should be averaged.
 This motivates us to consider the average of boundaries.
 }
 \label{fig:duality}
\end{figure}

During the past few years, boundary conformal field theories (BCFTs) have come to play an important role in the understanding of quantum gravity.
One main reason comes from a class of toy models (called the island model) where the black hole and a non-gravitational bath CFT are glued along the (asymptotic) boundary, which has been investigated to provide progress on the information paradox problem \cite{Penington2020,Almheiri2019,Almheiri2020}.
This model is closely related to BCFTs through the AdS/BCFT correspondence  \cite{Takayanagi2011,Fujita2011} and braneworld holography \cite{Karch2001,Randall1999,Randall1999a} (see FIG. \ref{fig:duality}).
This relation is applied to provide more insights into quantum gravity in many works (for example, see \cite{Ageev:2021ipd,Akal:2021foz,Chu:2021gdb,Bousso:2020kmy,Sully:2020pza,Akal2021,Akal2020,Miao2021,Geng2021, Geng2021a}).
Nevertheless, many aspects are still unclear.
One reason is a lack of knowledge about the mechanism of AdS/BCFT.
On this background, it would be important to address the following issue, ``how is the information about AdS gravity encoded in BCFT?
Particularly, we focus on a brane, which is a physical object that generalizes the notion of a point particle to higher dimensions.
One main goal of this article is to show how the brane physics is explained from the BCFT side.

It has been pointed out that in AdS/BCFT, one may have an unphysical solution where a brane has a self-intersection \cite{Cooper2019, Geng:2021iyq, Kawamoto2022, Bianchi2022}.
We also expect that non-trivial intersections (associated with non-zero tension) between two distinct branes are not allowed to exist.
One resolution may be obtained by just excluding such solutions, as proposed in \cite{Geng:2021iyq}.
In this article, we address this issue by the conformal bootstrap in BCFT and show that the brane intersection problem can be avoided by a black hole formation.

In this article,
we assume (and verify) that the BCFT dual to the semiclassical gravity can be obtained by the average of boundary conditions.
This concept is related to a particular question, ``which type of BCFT (right of FIG. \ref{fig:duality}) can have the island model picture (left of FIG. \ref{fig:duality})'' (a study in the same direction can be found in \cite{Suzuki2022,Kusuki2022,Anous2022,Izumi2022} ).
We propose that this is obtained by the average of boundary conditions.
We define the average in the same way as that introduced in a recent interesting paper \cite{Chandra2022} (see also \cite{Belin2021a}), which shows that the averaged CFT completely reproduces the on-shell action of gravity.
As a next step, one can naturally consider including boundaries in the averaged CFT.
As we can see in this article, one can define the averaged boundary condition in a reasonable way in this averaged CFT by developing the results in \cite{Kusuki2022}.

\section{Ensemble of Boundaries}

We are interested in two-dimensional CFTs with the Cardy boundaries \cite{Cardy2004},
which satisfy the following boundary condition for the stress tensor,
\begin{equation}\label{eq:T=T}
\left. \left(T(z) - \bar{T}(\bar{z})\right)\right|_{\text{bdy}} = 0.
\end{equation}
By the state/operator-like correspondence, boundaries can be mapped to states (called boundary states).
The Cardy boundary condition (\ref{eq:T=T}) can be re-expressed in terms of the boundary state $\ket{B^a}$ as
\begin{equation}
L_n \ket{B^a} = \bar{L}_{-n} \ket{B^a},
\end{equation}
where $L_n$ is the Virasoro generator and we label the boundary condition by the superscript $a$.
Any Cardy boundary states can be expressed in the following form,
\begin{equation}
\ket{B^a} = g^a \sum_p C^a_{p \mathbb{I}} \kett{p}.
\end{equation}
The states $\kett{p}$ is the Ishibashi state, which is defined as
\begin{equation}\label{eq:Ishibashi}
\ket{j}\rangle \equiv \sum_{N} \ket{j;N} \otimes   U \overline{\ket{j;N}},
\end{equation}
where $\ket{j;N}$ is a state in the Verma module $j$ labeled by $N$, and $U$ is an anti-unitary operator.
The $g$-function $g^a$ describes the disk partition function $\braket{\mathbb{I}}_a = g^a$.
Note that the coefficients have a nice interpretation, called the bulk-boundary OPE coefficients,
    \begin{equation}
        \phi_i (z) \sim \sum_I (2 \Im z)^{h_I-h_i-\bar{h}_i} C^a_{iI} \phi_I(\Re z) + \cdots.
    \end{equation}
Here, we denote primary fields in the bulk by $\phi$ with the lowercase letter $i$
and primary fields on the boundary by $\phi$ with the capital letter $I$.
The boundary primary field  $\phi_I (x)$ is a primary field with conformal dimension $h_I$, which lives only on the boundary.

In this language, the boundary condition is characterized by the coefficients $ C^a_{p \mathbb{I}}$.
We define an ensemble of boundaries
by treating the bulk-boundary OPE coefficients as Gaussian random variables with zero mean and variance given by the universal asymptotic formula \cite{Kusuki2022, Numasawa2022},
\begin{equation}\label{eq:universal}
\overline{C^a_{p\mathbb{I}} C^a_{p\mathbb{I}}  } = \pa{g^a}^{-2} S_{\mathbb{I} p}^{-1},
\end{equation}
where $S_{ij}$ is the modular-S matrix \cite{Zamolodchikov2001}.
We assume that $\overline{C^a_{p\mathbb{I}} } = 0$ only holds if $p \neq \mathbb{I}$.
In other words, we assume that the spectrum in the closed string sector has a unique normalizable vacuum state.
We comment on the relation to the island model.
As pointed out in \cite{Suzuki2022},
any one-point function on a disk vanishes in the island model.
This is not the case for general BCFTs.
The point is that our assumption $\overline{C^a_{p\mathbb{I}} } = 0$ completely reflects this property in BCFTs.
Note also that in the minimal gravity with no matter coupling on the brane, one can find the same property  $ C^a_{p \mathbb{I}}=0$ \cite{Fujita2011,Takayanagi2011} (see also \cite{Suzuki2022}).
For this reason, it is worth considering the bootstrap equation with the condition $\overline{C^a_{p\mathbb{I}} } = 0$.

The formula (\ref{eq:universal}) holds for any unitary compact CFTs if the conformal dimension $h_p$ is very large $h_p \gg c$.
Our requirement is that the universal formula (\ref{eq:universal}) holds for $h_p$ of order $c$.
We expect that this extension of the validity region can be applied to a class of CFTs, called holographic CFT,
which are CFTs with a large central charge and a sparse spectrum (see \cite{Hartman2014} and also \cite{Belin2017, Kraus2017a, Michel2019}).
It provides a natural ensemble:
black hole states with a Cardy spectrum, defect states, and Gaussian random OPE coefficients with mean zero and a particular variance (see  \cite{Chandra2022} for the explicit definition).
We define the ensemble of boundaries (\ref{eq:universal}) in that ensemble-averaged theory in the same manner.
Note that we do not use the randomness of the bulk-bulk-bulk OPE coefficients.
Note also that we focus only on the case where Gaussian ansatz can be applied to.
If one would like to investigate higher-momentum terms (related to multiboundary wormholes), one may need to modify the ansatz.

\section{Self-Intersection}

\newsavebox{\boxba}
\sbox{\boxba}{\includegraphics[width=40pt]{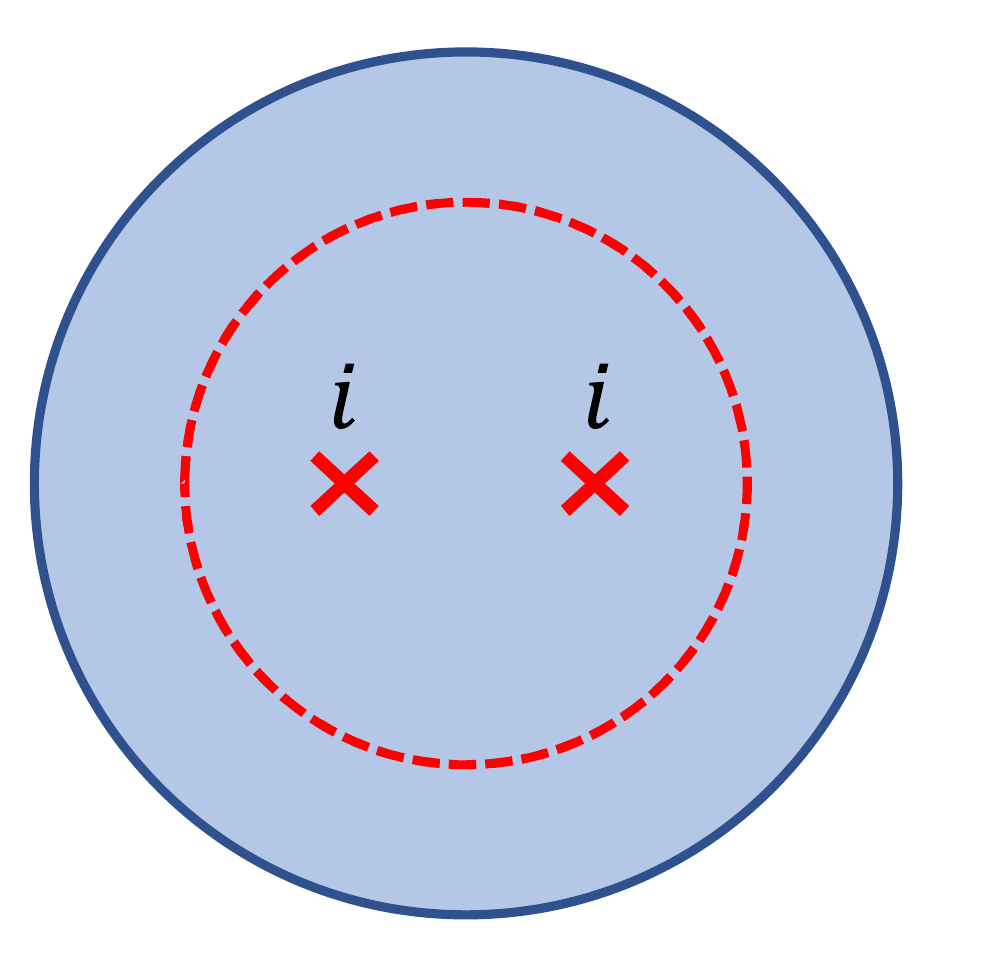}}
\newlength{\boxbaw}
\settowidth{\boxbaw}{\usebox{\boxba}} 

\newsavebox{\boxbb}
\sbox{\boxbb}{\includegraphics[width=40pt]{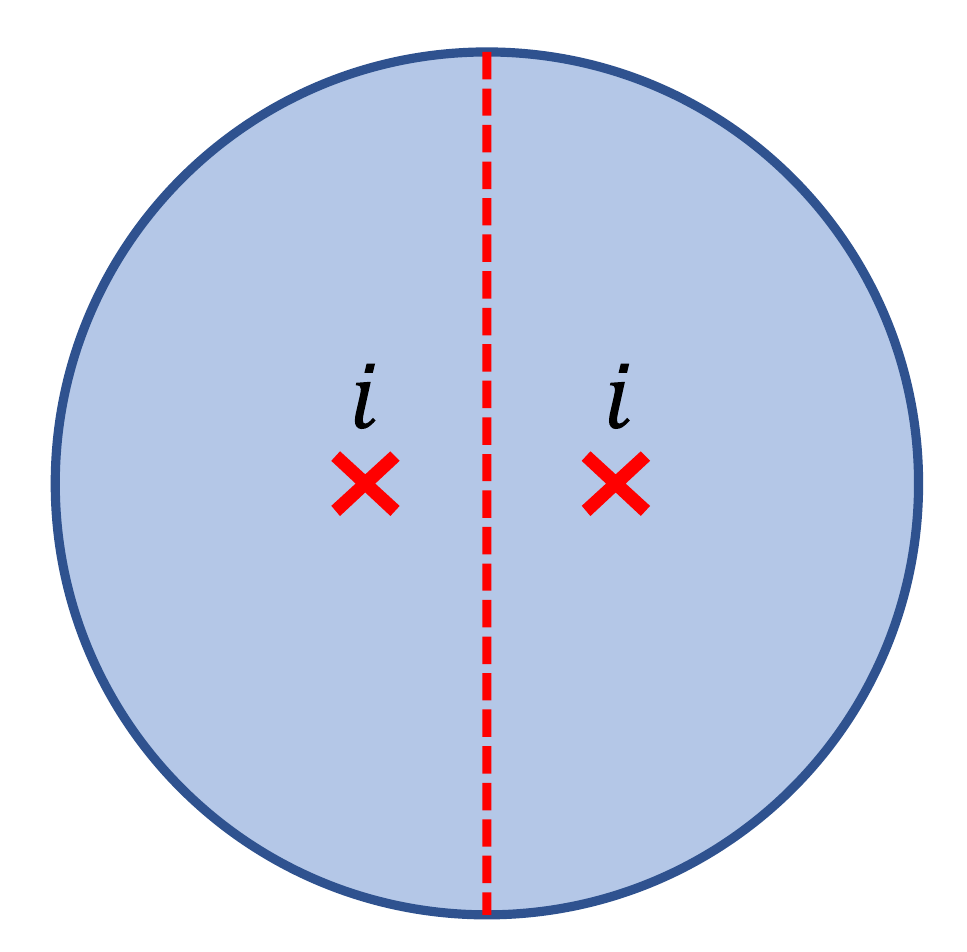}}
\newlength{\boxbbw}
\settowidth{\boxbbw}{\usebox{\boxbb}}

\begin{figure}[t]
 \begin{center}
  \includegraphics[width=10.0cm,clip]{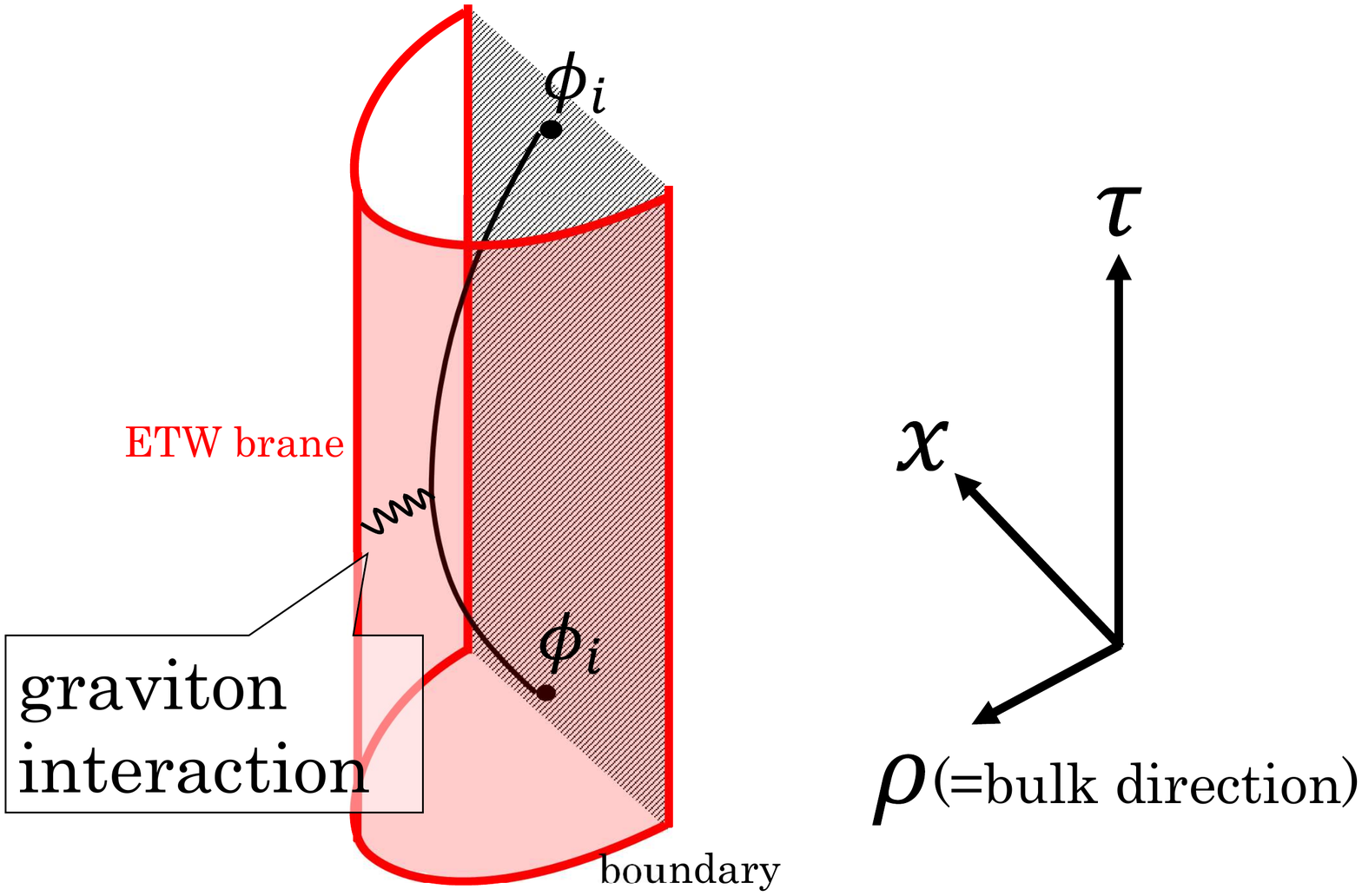}
 \end{center}
 \caption{
 The gravity dual of a bulk primary two-point function on a strip.
 This geometry is approximately time-independent since two bulk primaries $\phi_i$ are inserted at $\tau=\{ \infty, -\infty \}$.
 }
 \label{fig:setup}
\end{figure}

\begin{figure}[t]
 \begin{center}
  \includegraphics[width=8.0cm,clip]{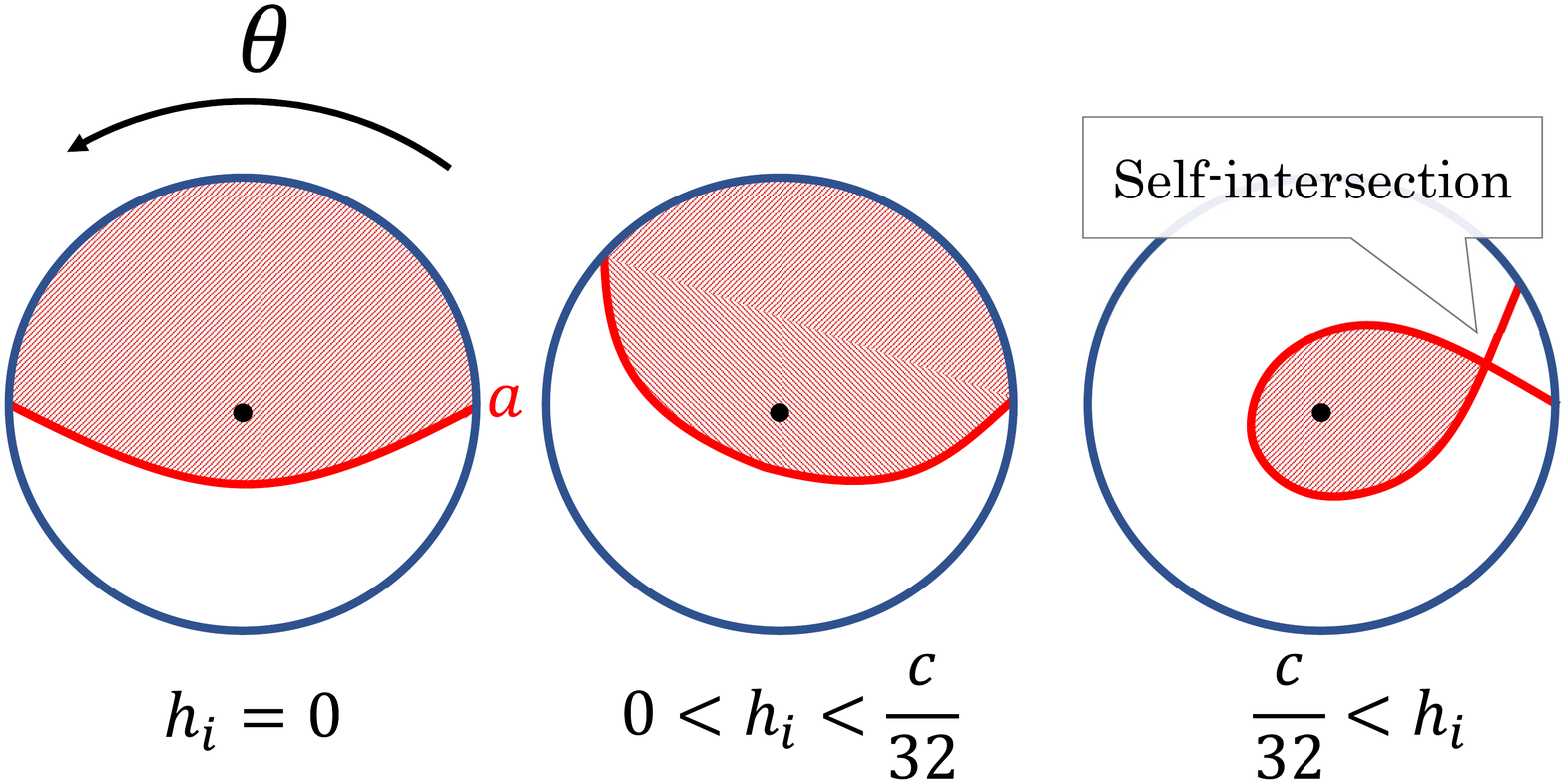}
 \end{center}
 \caption{The profile of the ETW brane in geometry with a conical defect. 
 The red shaded region is the bulk region and the red line describes the ETW brane.
 This conical defect is created by putting bulk primaries with the conformal dimension $h_i=\bar{h}_i$ on the asymptotic boundary.
 For $h_i=0$, the brane ends on $\theta=\{0,\pi\}$ (see the left picture).
 For $h_i > \fr{c}{32}$, the brane has a self-intersection (see the right picture).
 }
 \label{fig:self}
\end{figure}

In the holographic dual of BCFT, there is an interesting object, a brane (see the center of FIG. \ref{fig:duality}).
The existence of branes brings up various interesting questions.
Here, we will address one important issue about the brane.

As pointed out in \cite{Geng:2021iyq}, if one puts two bulk primaries $\phi_i$ at $\tau=\{\infty, -\infty\}$ on a strip (see our setup, FIG.\ref{fig:setup}),
the gravity dual can have a self-intersecting brane if $h_i > \fr{c}{32}$ where $h_i=\bar{h}_i$ is the conformal dimension of the bulk primary (see FIG.\ref{fig:self}).
This is because a particle with mass $\Delta_i=h_i+\bar{h}_i$ produces a deficit angle, $\delta\theta= 2\pi \pa{1-\sqrt{1-\fr{24h_i}{c}}}$
and the brane intersects with itself if $\delta\theta>\pi$, which provides the self-intersection bound $h_i \leq \fr{c}{32}$.
The authors in \cite{Geng:2021iyq} have proposed that primary operators in $(\fr{c}{32}, \fr{c}{24} )$ should be excluded in the holographic CFT.
The value $\fr{c}{24}$ comes from the BTZ threshold
\footnote{
There is an analog of the Hawking-Page transition in AdS/BCFT \cite{Fujita2011, Takayanagi2011}.
}.
If so, the self-interaction can be avoided by a black hole formation above the BTZ threshold.
Here, we will show that the self-intersection brane actually does not appear in the averaged boundaries even below $\fr{c}{24}$.

In gravity with the ETW branes, particles are interacting with not only other particles but also branes.
With the doubling trick \cite{Cardy2004}, we can translate it into the interaction with its mirror.
This interaction makes the binding energy of the one-particle state (or the two-particle state with its mirror) non-trivial.
We identify this energy by using the conformal bootstrap.
Our interest is a two-point function on a disk.
For this correlation function, we have the following two choices of how to cut
\footnote{
We mean cut by inserting the identity operator $\Sigma_p \ket{p} \bra{p}$.
},
\begin{equation}
\parbox{\boxbaw}{\usebox{\boxba}} = \parbox{\boxbbw}{\usebox{\boxbb}}.
\end{equation}
The corresponding bootstrap equation is given by (see the details of the bootstrap equations in BCFTs in \cite{Kusuki2022}),
\begin{equation}
\sum_{p} C_{iip}C^a_{p\mathbb{I}} \ca{F}^{ii}_{ii}(h_p|z) 
=
\sum_{P}\pa{C^a_{iP}}^2 \ca{F}^{ii}_{ii}(h_P|1-z),
\end{equation}
where $C_{iip}$ is bulk-bulk-bulk OPE coefficients and $\ca{F}^{ii}_{ii} (h_p|z)$ is the Virasoro conformal block.
We define the cross ratio $z \equiv \fr{z_{12}{z_{34}}}{z_{13}z_{24}}$ by the insertion points of two bulk primary operators $z_1$ and $z_2$ with the relation $z_3=z_2^*$, $z_4=z_1^*$.
The sum on the left-hand side runs over primaries in the closed string sector and that on the right-hand side runs over primaries in the open string sector.
The energy of the one-particle state is given by the lowest conformal dimension in the right-hand side.
For the ensemble-average, we have $\overline{C^a_{p\mathbb{I}}}=\delta_{p\mathbb{I}}$ and then the bootstrap equation is simplified as
\footnote{
We may be able to give the same approximation by $\overline{C}_{iip}=0$ and no correlation between $C_{iip}$ and $C_{p\mathbb{I}}^a$.
One can also realize $C^a_{p \mathbb{I}}=0$ by setting $p\neq \bar{p}$.
We give further comments on this possibility in \cite{Kusuki}.
}
\footnote{
There might be a case where we cannot approximate it by the vacuum. We will discuss the details in \cite{Kusuki}.
}
\begin{equation}\label{eq:two}
\ca{F}^{ii}_{ii}(0|z) 
=
\int \dd \alpha_P \rho^{aa}( \alpha_P ) \overline{\pa{C^a_{iP}}^2} \ca{F}^{ii}_{ii}(h_P|1-z).
\end{equation}
For convenience, 
we introduce the notation usually found in the Liouville CFT,
\begin{equation}
c=1+6Q^2, \ \ \ \ \ Q=b+\fr{1}{b}, \ \ \ \ \ h_i=\a_i(Q-\a_i).
\end{equation}
We also define the degeneracy of primary states $\rho (\a)$ as
\begin{equation}
\rho (\a)=\sum_p D_p \delta(\a-\a_p) ,
\end{equation}
where the function $D_p$ denotes the degeneracy of primary operators with the Liouville momentum $\alpha_p$.
The bootstrap equation (\ref{eq:two}) can be solved in the following way.
The left-hand side can be expanded in the same basis as the right-hand side by using the fusion transformation \cite{Teschner2001,Ponsot1999,Teschner2003,Ponsot2001}.
Following the notation in \cite{Kusuki2019, Kusuki2019a}, the fusion transformation is expressed by
\footnote{
This transformation is usually expressed only by an integral,
but as expressed here, the precise form has an additional summation.
The summation comes from the poles of the fusion matrix \cite{Teschner2001}.
One can see the details in, for example, Appendix A of \cite{Kusuki2019a}.
}
\begin{equation}\label{eq:fusion}
\begin{aligned}
&\ca{F}^{AA}_{BB}(0|z)\\
&=
 \sum_{\substack{\a_{n,m}<\fr{Q}{2} \\ n,m \in \bb{Z}_{\geq0}}}\ \text{Res}  \biggl( \biggr.   -2\pi i 
  {\bold F}_{0, \a_t} 
   \left[
    \begin{array}{cc}
    \a_A   & \a_A  \\
     \a_B  &   \a_B\\
    \end{array}
  \right] \\
  &\ca{F}^{AB}_{AB}(h_{\a_t}|1-z);\a_t=\a_{n,m}  \biggl. \biggr)     \\
&+
\int_{\fr{Q}{2}+0}^{\fr{Q}{2}+i \infty} \dd \a_t {\bold F}_{0, \a_t} 
   \left[
    \begin{array}{cc}
    \a_A   & \a_A  \\
     \a_B  &   \a_B\\
    \end{array}
  \right]
  \ca{F}^{AB}_{AB}(h_{\a_t}|1-z),
\end{aligned}
\end{equation}
where $\a_{n,m}\equiv\a_A+\a_B+mb+nb^{-1}$ and the kernel $ {\bold F}_{\a_s, \a_t} $ denotes the fusion matrix (or crossing matrix), whose closed form is derived in \cite{Teschner2001}.
Comparing the coefficients in the equation (\ref{eq:two}), we can identify the bulk-boundary OPE coefficients by the fusion matrix.
Note that the CFT whose OPE coefficients are given by the fusion matrix is called the Virasoro mean-field theory (VMFT) \cite{ Collier2019}.
The lowest energy is given by $\alpha_P=2\alpha_i$, which corresponds to the mass of the excited geometry.
It implies that when the conformal dimension of the external operator satisfies $h_i >\fr{c}{32}$ (i.e., $\Re \alpha_i>\fr{Q}{4}$), the one-particle state (interacting with the brane) has the energy $h_P > \fr{c}{24}$ (i.e., $\Re\alpha_P\geq \fr{Q}{2}$).
Therefore, a black hole is formed in this geometry and consequently, the brane can avoid the self-intersection.

This non-trivial modification of the black hole formation threshold $\fr{c}{24} \to \fr{c}{32}$ for the boundary primary conformal dimension comes from the existence of the binding energy. The detailed explanation is as follows.
Unlike higher-dimensional AdS${}_d$ ($d\geq4$), the gravitational interactions in AdS${}_3$ create a deficit angle, which can be detected even at infinite separation.
As a result, even if non-trivial (matter) interactions are turned off like (\ref{eq:two}), the gravitational interaction non-trivially contributes.
This observation suggests that the gravity calculation in AdS/BCFT requires careful treatment.
We will show a detailed analysis in \cite{Kusuki}, where we give a completely consistent result about the ADM mass from the gravity side.
Note that this long-range universal interaction can also be found in the spectrum of the large-spin two-particle state as a non-trivial binding energy \cite{Fitzpatrick2014, Kusuki2019a, Collier2019}.
Actually, in \cite{Kusuki2019a, Collier2019}, one can also see the value $\fr{c-1}{32}$ as the black hole formation bound for a two-particle state with a large spin.
This coincidence comes from the same vacuum block approximation of the bootstrap equation.
The large-spin spectrum is identified by the vacuum block approximation of the bootstrap equation in the light-cone limit.
In fact, this approximated bootstrap equation is almost the same as (\ref{eq:two}) even though this comes from not the light-cone limit but the averaging.
That is the reason why we found the value $\fr{c-1}{32}$ in these different contexts.

\section{Intersection with Another Brane}

Let us consider a partition function on a cylinder with the boundary conditions ($a$,$b$).
There are two candidates for the brane configuration in the gravity dual (see \cite{Takayanagi2011,Fujita2011}).
The one is connected and the other is disconnected.
If we consider two distinct boundary conditions $a\neq b$,
only the disconnected phase should be allowed.
In this case, the branes can naively have intersections.
We will see whether this type of intersection can be avoided in our averaged BCFT by using the conformal bootstrap.

The conformal bootstrap equation for a cylinder partition function with the boundary condition $(a,b)$ is given by
\begin{equation}\label{eq:cylinder}
\begin{aligned}
g^a g^b
\sum_{p }
C^a_{p\mathbb{I}}  C^b_{p\mathbb{I}} 
\chi_p(\tau)
=
\sum_{P }
\chi_P \pa{-\fr{1}{\tau}},
\end{aligned}
\end{equation}
where $ \chi_{p}(\tau)$ is the Virasoro character.
Averaging over boundaries leads to
\begin{equation}\label{eq:vari}
\begin{aligned}
\overline{C^a_{p\mathbb{I}}  C^b_{p\mathbb{I}} } 
&=\left\{
    \begin{array}{ll}
     \delta_{ab} \pa{g^a g^b S_{\mathbb{I}p}}^{-1}  ,& \text{if } p\neq \mathbb{I}  ,\\
      1 ,& \text{if }  p =  \mathbb{I}  .\\
    \end{array}
  \right.\\
\end{aligned}
\end{equation}
Thus we obtain the following averaged conformal bootstrap equation for $a \neq b$,
\begin{equation}
\begin{aligned}
g^a g^b
\chi_\mathbb{I}(\tau)
=
\int\dd \a_P \ 
\rho^{ab} (\a_P)
\chi_P \pa{-\fr{1}{\tau}},
\end{aligned}
\end{equation}
where $\rho^{ab} (\a)$ is the density of primary states in the open string sector with two boundary conditions $(a,b)$.
The left-hand side can be expanded in the dual channel basis as \cite{Zamolodchikov2001} 
\begin{equation}
\chi_\mathbb{I}(\tau)
=
\int^{\fr{Q}{2}+i\infty}_{\fr{Q}{2}-i\infty}  \dd \a_P \ 
S_{\mathbb{I}P}
\chi_P \pa{-\fr{1}{\tau}}.
\end{equation}
The closed form of the modular-$S$ matrix is given in \cite{Zamolodchikov2001}.
As a result, we obtain the relation,
\begin{equation}\label{eq:rhobb}
\begin{aligned}
\rho^{ab} (\a_P)
&=\left\{
    \begin{array}{ll}
    g^a g^b   S_{\mathbb{I}P}    ,& \text{if } \alpha_P=\fr{Q}{2}+i\mathbb{R}  ,\\
    0   ,& \text{otherwise }   .\\
    \end{array}
  \right.\\
\end{aligned}
\end{equation}
It implies that the spectrum in the open string sector with $a \neq b$ has only black hole states $h_P\geq\fr{c}{24}$.

\begin{figure}[t]
 \begin{center}
  \includegraphics[width=8.0cm,clip]{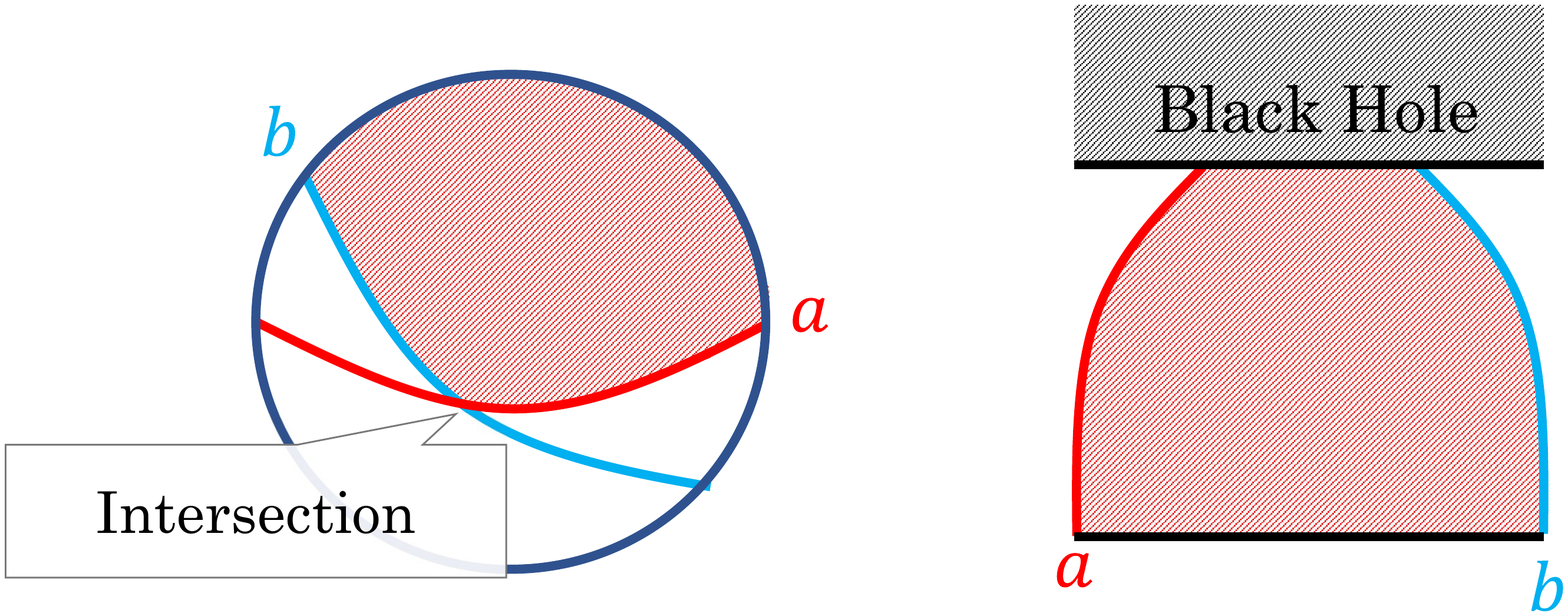}
 \end{center}
 \caption{The profile of the gravity with two different ETW branes. 
 The red region describes the bulk region. This is sandwiched by two ETW branes.
 The configuration shown in the left picture does not appear in our theory.
 The intersection can be avoided by a black hole formation as shown in the right picture.}
 \label{fig:inter}
\end{figure}

Let us move on to the gravity interpretation.
The holographic dual of the strip with two different boundaries $a\neq b$ may have an intersection (see the left of FIG.\ref{fig:inter}).
This configuration may not be reasonable.
Our result (\ref{eq:rhobb}) suggests that in the semiclassical gravity (or averaged BCFT),
the intersection can be avoided by a black hole formation (see the right of FIG.\ref{fig:inter}).
Note that this result is consistent with \cite{Miyaji2021}
\footnote{
The authors in \cite{Miyaji2021} propose that the boundary state behaves like a random state in the holographic CFT.
But their definition of the randomness is quite different from ours.
}
.

The spectrum (\ref{eq:rhobb}) does not include the vacuum state.
We would like to comment that this is not the case for the identical boundary condition $a=b$.
In this case, the left-hand side of (\ref{eq:cylinder}) includes the contribution,
\begin{equation}
\int\dd \a_p \ 
S_{\mathbb{I}p}
\chi_p \pa{\tau}
=
\chi_{\mathbb{I}} \pa{-\fr{1}{\tau}},
\end{equation}
where we use the upper equation of (\ref{eq:vari}) and the universal formula for the density of bulk primary states $\rho(\a_p,\a_{\bar{p}}) = S_{\mathbb{I}p}  S_{\mathbb{I}\bar{p}} $ (see \cite{Kusuki2022}).
Hence, the spectrum in the open string sector $\rho^{aa}(\alpha_P)$ includes the vacuum.
We would like to emphasize that the open string spectrum with the identical boundary condition $\rho^{aa}$ should include the vacuum.
Our averaging provides a natural realization of the boundary satisfying both two expected properties, the existence of the vacuum in $\rho^{aa}$ and no intersections.
It is worth noting that if one just assumes the condition $C_{p\mathbb{I}}^a=0$ (without averaging), one finds the absence of the vacuum in the open string spectrum with the identical boundary condition.

\section{Wormholes}

In this section, we consider the average of a product of correlation functions, which produces wormholes on the gravity side.
Although it would be interesting to give a comprehensive analysis of wormholes in AdS/BCFT,
we only give one example since calculations of other examples could be straightforward as in \cite{Chandra2022}.

Let us consider a product of two bulk two-point functions on a disk,
\begin{equation}
\begin{aligned}
\braket{\phi_i \phi_i}_{\text{disk}} \braket{\phi_i \phi_i}_{\text{disk}}
=
\sum_{p } \pa{C_{iip}}^2   \pa{C^a_{p\mathbb{I}}}^2  \ca{F}^{ii}_{ii}(h_p|z) \ca{F}^{ii}_{ii}(h_p|z').
\end{aligned}
\end{equation}
If we take the average, we find not only the vacuum contribution but also the following contributions in the large $c$ limit,
\begin{equation}
\begin{aligned}
\sum_{p } \overline{\pa{C_{iip}}^2}  \overline{ \pa{C^a_{p\mathbb{I}}}^2 } \ca{F}^{ii}_{ii}(h_p|z) \ca{F}^{ii}_{ii}(h_p|z') 
=G_{iiii}^L(z,z')
\end{aligned}
\end{equation}
where we define the average of the bulk-bulk-bulk OPE coefficients by Gaussian random OPE coefficients with mean zero and variance,
$
  {\bold F}_{0, \a_p} 
   \left[
    \begin{array}{cc}
    \a_i   & \a_i  \\
     \a_i  &   \a_i\\
    \end{array}
  \right]
 /
  S_{\mathbb{I} p}
$
(see \cite{Chandra2022} for details).
The function $G_{iiii}^L(z,\bar{z})$ is the Liouville four-point function.
The relation to the Liouville CFT comes from the relation between the fusion matrix and the DOZZ OPE coefficients \cite{Collier2020}.

This extra contribution represented by the Liouville correlator corresponds to the wormhole contribution on the gravity side.
In this article, we do not explain it in detail because the mechanism is completely the same as that explained in \cite{Chandra2022}.

Here we will give a brief comment on the connection to the island model. The details will be found in \cite{Kusuki}.
We would like to argue that this model can be interpreted as the BCFT dual of the island model.
Through the AdS/CFT, one may be able to map the JT gravity in the island model to a boundary (see the upper equation in FIG. \ref{fig:duality}).
This boundary corresponds to the averaged SYK model \footnote{
This SYK model should be modified since it is non-trivially coupled to the non-gravitational bath CFT.
}.
This consideration motivates us to consider an average of boundary conditions.
On this background, our model may realize the simplest model that captures the essential properties of the island model.
At first glance, the correlation function with two twist operators on a disk seems to have only the connected phase (the vacuum block contribution in a bulk channel) since $\overline{C_{\sigma_n \bar{\sigma}_n p}} \overline{C^a_{p\mathbb{I}}}=0$ if $p\neq \mathbb{I}$.
In other words, there seems to be no island contribution to the entanglement entropy.
 In fact, this is not true because there are wormhole contributions if taking the boundary average.
If one considers the $n$-replica sheet with boundaries, one finds that each replica sheet is connected via the wormhole between the boundaries after taking the average.
An interesting future direction would be relating these two models in an explicit way.

So far, we think of the boundary as a theory (i.e., SYK model), or equivalently, the boundary average as a theory average.
We would like to comment that
we can also think of the boundary average as a state average.
If the bulk-boundary OPE coefficients follow the Gaussian ensemble,
the following state is almost the same as the thermal pure quantum (TPQ) state \cite{Sugiura2012,Sugiura2013},
\begin{equation}
\ket{TPQ} = \ex{-\fr{\beta}{2}H}\ket{B}.
\end{equation}
It has been proposed that the state averaging effectively provides spatial wormholes \cite{Freivogel2021, Goto2021} (see also \cite{Pollack2020, Altland2021a, Altland2021, Belin2021a} for the connection between the wormholes and the state averaging).
Thus, one can relate the spatial wormholes to the boundary average.
The point is that the theory averaging and the state averaging can be obtained from the same averaging of boundaries.
Therefore, the averaging over boundaries may have the potential to provide a clear understanding of the relation between the state and theory averaging.

\section{Discussions}
There are many interesting future works.
We put some of them here.
It would be interesting to give an explicit example of the boundary averaging.
One simple realization may be obtained in the averaged Narain CFT \cite{Maloney2020,AfkhamiJeddi2021}.
The compactified boson can have Dirichlet, Neumann, and Friedan \cite{Janik2001} boundary conditions
\footnote{
It has not been proven that this is the full classification of boundaries. 
}
.
We may be able to consider the average of these boundary conditions.
We hope to see further investigations in this direction in the future paper.

Besides bulk-bulk-bulk OPE coefficients, BCFTs have bulk-boundary OPE coefficients and boundary-boundary-boundary OPE coefficients.
The universal formula for these additional ingredients has been provided in \cite{Kusuki2022, Numasawa2022}.
Based on these results, one can define the averaged BCFT (partly defined in this article by averaging the bulk-boundary OPE coefficients).
An interesting future direction would be checking whether the matching between the averaged CFT calculation and the gravity calculation \cite{Chandra2022} also works in BCFTs in an explicit way.

We proposed that the boundary averaging provides a simple version of the island model.
The point is that the calculation in the averaged BCFT may be easier than in the original island model which requires solving the conformal welding problem \cite{Almheiri2020a}.
Thus, we may be able to address more complicated but more interesting issues through this averaged BCFT.

\section*{Acknowledgments}
We are grateful to Nathan Benjamin, Cyuan-Han Chang, Scott Collier, Diptarka Das, Masamichi Miyaji, Takato Mori, Sridip Pal, David Simmons-Duffin, and Zixia Wei for useful discussions and comments.
YK is supported by Burke Fellowship (Brinson Postdoctoral Fellowship).

\bibliography{main}

\begin{thebibliography}{65}
\expandafter\ifx\csname natexlab\endcsname\relax\def\natexlab#1{#1}\fi
\expandafter\ifx\csname bibnamefont\endcsname\relax
  \def\bibnamefont#1{#1}\fi
\expandafter\ifx\csname bibfnamefont\endcsname\relax
  \def\bibfnamefont#1{#1}\fi
\expandafter\ifx\csname citenamefont\endcsname\relax
  \def\citenamefont#1{#1}\fi
\expandafter\ifx\csname url\endcsname\relax
  \def\url#1{\texttt{#1}}\fi
\expandafter\ifx\csname urlprefix\endcsname\relax\def\urlprefix{URL }\fi
\providecommand{\bibinfo}[2]{#2}
\providecommand{\eprint}[2][]{\url{#2}}

\bibitem[{\citenamefont{Penington}(2020)}]{Penington2020}
\bibinfo{author}{\bibfnamefont{G.}~\bibnamefont{Penington}},
  \bibinfo{journal}{JHEP} \textbf{\bibinfo{volume}{09}}, \bibinfo{pages}{002}
  (\bibinfo{year}{2020}), \eprint{1905.08255}.

\bibitem[{\citenamefont{Almheiri et~al.}(2019)\citenamefont{Almheiri,
  Engelhardt, Marolf, and Maxfield}}]{Almheiri2019}
\bibinfo{author}{\bibfnamefont{A.}~\bibnamefont{Almheiri}},
  \bibinfo{author}{\bibfnamefont{N.}~\bibnamefont{Engelhardt}},
  \bibinfo{author}{\bibfnamefont{D.}~\bibnamefont{Marolf}}, \bibnamefont{and}
  \bibinfo{author}{\bibfnamefont{H.}~\bibnamefont{Maxfield}},
  \bibinfo{journal}{JHEP} \textbf{\bibinfo{volume}{12}}, \bibinfo{pages}{063}
  (\bibinfo{year}{2019}), \eprint{1905.08762}.

\bibitem[{\citenamefont{Almheiri
  et~al.}(2020{\natexlab{a}})\citenamefont{Almheiri, Mahajan, Maldacena, and
  Zhao}}]{Almheiri2020}
\bibinfo{author}{\bibfnamefont{A.}~\bibnamefont{Almheiri}},
  \bibinfo{author}{\bibfnamefont{R.}~\bibnamefont{Mahajan}},
  \bibinfo{author}{\bibfnamefont{J.}~\bibnamefont{Maldacena}},
  \bibnamefont{and} \bibinfo{author}{\bibfnamefont{Y.}~\bibnamefont{Zhao}},
  \bibinfo{journal}{JHEP} \textbf{\bibinfo{volume}{03}}, \bibinfo{pages}{149}
  (\bibinfo{year}{2020}{\natexlab{a}}), \eprint{1908.10996}.

\bibitem[{\citenamefont{Takayanagi}(2011)}]{Takayanagi2011}
\bibinfo{author}{\bibfnamefont{T.}~\bibnamefont{Takayanagi}},
  \bibinfo{journal}{Phys. Rev. Lett.} \textbf{\bibinfo{volume}{107}},
  \bibinfo{pages}{101602} (\bibinfo{year}{2011}), \eprint{1105.5165}.

\bibitem[{\citenamefont{Fujita et~al.}(2011)\citenamefont{Fujita, Takayanagi,
  and Tonni}}]{Fujita2011}
\bibinfo{author}{\bibfnamefont{M.}~\bibnamefont{Fujita}},
  \bibinfo{author}{\bibfnamefont{T.}~\bibnamefont{Takayanagi}},
  \bibnamefont{and} \bibinfo{author}{\bibfnamefont{E.}~\bibnamefont{Tonni}},
  \bibinfo{journal}{JHEP} \textbf{\bibinfo{volume}{11}}, \bibinfo{pages}{043}
  (\bibinfo{year}{2011}), \eprint{1108.5152}.

\bibitem[{\citenamefont{Karch and Randall}(2001)}]{Karch2001}
\bibinfo{author}{\bibfnamefont{A.}~\bibnamefont{Karch}} \bibnamefont{and}
  \bibinfo{author}{\bibfnamefont{L.}~\bibnamefont{Randall}},
  \bibinfo{journal}{JHEP} \textbf{\bibinfo{volume}{05}}, \bibinfo{pages}{008}
  (\bibinfo{year}{2001}), \eprint{hep-th/0011156}.

\bibitem[{\citenamefont{Randall and Sundrum}(1999{\natexlab{a}})}]{Randall1999}
\bibinfo{author}{\bibfnamefont{L.}~\bibnamefont{Randall}} \bibnamefont{and}
  \bibinfo{author}{\bibfnamefont{R.}~\bibnamefont{Sundrum}},
  \bibinfo{journal}{Phys. Rev. Lett.} \textbf{\bibinfo{volume}{83}},
  \bibinfo{pages}{3370} (\bibinfo{year}{1999}{\natexlab{a}}),
  \eprint{hep-ph/9905221}.

\bibitem[{\citenamefont{Randall and
  Sundrum}(1999{\natexlab{b}})}]{Randall1999a}
\bibinfo{author}{\bibfnamefont{L.}~\bibnamefont{Randall}} \bibnamefont{and}
  \bibinfo{author}{\bibfnamefont{R.}~\bibnamefont{Sundrum}},
  \bibinfo{journal}{Phys. Rev. Lett.} \textbf{\bibinfo{volume}{83}},
  \bibinfo{pages}{4690} (\bibinfo{year}{1999}{\natexlab{b}}),
  \eprint{hep-th/9906064}.

\bibitem[{\citenamefont{Ageev}(2021)}]{Ageev:2021ipd}
\bibinfo{author}{\bibfnamefont{D.~S.} \bibnamefont{Ageev}}
  (\bibinfo{year}{2021}), \eprint{2107.09083}.

\bibitem[{\citenamefont{Akal et~al.}(2021{\natexlab{a}})\citenamefont{Akal,
  Kusuki, Shiba, Takayanagi, and Wei}}]{Akal:2021foz}
\bibinfo{author}{\bibfnamefont{I.}~\bibnamefont{Akal}},
  \bibinfo{author}{\bibfnamefont{Y.}~\bibnamefont{Kusuki}},
  \bibinfo{author}{\bibfnamefont{N.}~\bibnamefont{Shiba}},
  \bibinfo{author}{\bibfnamefont{T.}~\bibnamefont{Takayanagi}},
  \bibnamefont{and} \bibinfo{author}{\bibfnamefont{Z.}~\bibnamefont{Wei}},
  \bibinfo{journal}{Class. Quant. Grav.} \textbf{\bibinfo{volume}{38}},
  \bibinfo{pages}{224001} (\bibinfo{year}{2021}{\natexlab{a}}),
  \eprint{2106.11179}.

\bibitem[{\citenamefont{Chu et~al.}(2021)\citenamefont{Chu, Deng, and
  Zhou}}]{Chu:2021gdb}
\bibinfo{author}{\bibfnamefont{J.}~\bibnamefont{Chu}},
  \bibinfo{author}{\bibfnamefont{F.}~\bibnamefont{Deng}}, \bibnamefont{and}
  \bibinfo{author}{\bibfnamefont{Y.}~\bibnamefont{Zhou}},
  \bibinfo{journal}{JHEP} \textbf{\bibinfo{volume}{10}}, \bibinfo{pages}{149}
  (\bibinfo{year}{2021}), \eprint{2105.09106}.

\bibitem[{\citenamefont{Bousso and Wildenhain}(2020)}]{Bousso:2020kmy}
\bibinfo{author}{\bibfnamefont{R.}~\bibnamefont{Bousso}} \bibnamefont{and}
  \bibinfo{author}{\bibfnamefont{E.}~\bibnamefont{Wildenhain}},
  \bibinfo{journal}{Phys. Rev. D} \textbf{\bibinfo{volume}{102}},
  \bibinfo{pages}{066005} (\bibinfo{year}{2020}), \eprint{2006.16289}.

\bibitem[{\citenamefont{Sully et~al.}(2021)\citenamefont{Sully, Raamsdonk, and
  Wakeham}}]{Sully:2020pza}
\bibinfo{author}{\bibfnamefont{J.}~\bibnamefont{Sully}},
  \bibinfo{author}{\bibfnamefont{M.~V.} \bibnamefont{Raamsdonk}},
  \bibnamefont{and} \bibinfo{author}{\bibfnamefont{D.}~\bibnamefont{Wakeham}},
  \bibinfo{journal}{JHEP} \textbf{\bibinfo{volume}{03}}, \bibinfo{pages}{167}
  (\bibinfo{year}{2021}), \eprint{2004.13088}.

\bibitem[{\citenamefont{Akal et~al.}(2021{\natexlab{b}})\citenamefont{Akal,
  Kusuki, Shiba, Takayanagi, and Wei}}]{Akal2021}
\bibinfo{author}{\bibfnamefont{I.}~\bibnamefont{Akal}},
  \bibinfo{author}{\bibfnamefont{Y.}~\bibnamefont{Kusuki}},
  \bibinfo{author}{\bibfnamefont{N.}~\bibnamefont{Shiba}},
  \bibinfo{author}{\bibfnamefont{T.}~\bibnamefont{Takayanagi}},
  \bibnamefont{and} \bibinfo{author}{\bibfnamefont{Z.}~\bibnamefont{Wei}},
  \bibinfo{journal}{Phys. Rev. Lett.} \textbf{\bibinfo{volume}{126}},
  \bibinfo{pages}{061604} (\bibinfo{year}{2021}{\natexlab{b}}),
  \eprint{2011.12005}.

\bibitem[{\citenamefont{Akal et~al.}(2020)\citenamefont{Akal, Kusuki,
  Takayanagi, and Wei}}]{Akal2020}
\bibinfo{author}{\bibfnamefont{I.}~\bibnamefont{Akal}},
  \bibinfo{author}{\bibfnamefont{Y.}~\bibnamefont{Kusuki}},
  \bibinfo{author}{\bibfnamefont{T.}~\bibnamefont{Takayanagi}},
  \bibnamefont{and} \bibinfo{author}{\bibfnamefont{Z.}~\bibnamefont{Wei}},
  \bibinfo{journal}{Phys. Rev. D} \textbf{\bibinfo{volume}{102}},
  \bibinfo{pages}{126007} (\bibinfo{year}{2020}), \eprint{2007.06800}.

\bibitem[{\citenamefont{Miao}(2021)}]{Miao2021}
\bibinfo{author}{\bibfnamefont{R.-X.} \bibnamefont{Miao}},
  \bibinfo{journal}{JHEP} \textbf{\bibinfo{volume}{01}}, \bibinfo{pages}{150}
  (\bibinfo{year}{2021}), \eprint{2009.06263}.

\bibitem[{\citenamefont{Geng et~al.}(2021{\natexlab{a}})\citenamefont{Geng,
  Karch, Perez-Pardavila, Raju, Randall, Riojas, and Shashi}}]{Geng2021}
\bibinfo{author}{\bibfnamefont{H.}~\bibnamefont{Geng}},
  \bibinfo{author}{\bibfnamefont{A.}~\bibnamefont{Karch}},
  \bibinfo{author}{\bibfnamefont{C.}~\bibnamefont{Perez-Pardavila}},
  \bibinfo{author}{\bibfnamefont{S.}~\bibnamefont{Raju}},
  \bibinfo{author}{\bibfnamefont{L.}~\bibnamefont{Randall}},
  \bibinfo{author}{\bibfnamefont{M.}~\bibnamefont{Riojas}}, \bibnamefont{and}
  \bibinfo{author}{\bibfnamefont{S.}~\bibnamefont{Shashi}},
  \bibinfo{journal}{SciPost Phys.} \textbf{\bibinfo{volume}{10}},
  \bibinfo{pages}{103} (\bibinfo{year}{2021}{\natexlab{a}}),
  \eprint{2012.04671}.

\bibitem[{\citenamefont{Geng et~al.}(2021{\natexlab{b}})\citenamefont{Geng,
  Karch, Perez-Pardavila, Raju, Randall, Riojas, and Shashi}}]{Geng2021a}
\bibinfo{author}{\bibfnamefont{H.}~\bibnamefont{Geng}},
  \bibinfo{author}{\bibfnamefont{A.}~\bibnamefont{Karch}},
  \bibinfo{author}{\bibfnamefont{C.}~\bibnamefont{Perez-Pardavila}},
  \bibinfo{author}{\bibfnamefont{S.}~\bibnamefont{Raju}},
  \bibinfo{author}{\bibfnamefont{L.}~\bibnamefont{Randall}},
  \bibinfo{author}{\bibfnamefont{M.}~\bibnamefont{Riojas}}, \bibnamefont{and}
  \bibinfo{author}{\bibfnamefont{S.}~\bibnamefont{Shashi}}
  (\bibinfo{year}{2021}{\natexlab{b}}), \eprint{2112.09132}.

\bibitem[{\citenamefont{Cooper et~al.}(2019)\citenamefont{Cooper, Rozali,
  Swingle, Van~Raamsdonk, Waddell, and Wakeham}}]{Cooper2019}
\bibinfo{author}{\bibfnamefont{S.}~\bibnamefont{Cooper}},
  \bibinfo{author}{\bibfnamefont{M.}~\bibnamefont{Rozali}},
  \bibinfo{author}{\bibfnamefont{B.}~\bibnamefont{Swingle}},
  \bibinfo{author}{\bibfnamefont{M.}~\bibnamefont{Van~Raamsdonk}},
  \bibinfo{author}{\bibfnamefont{C.}~\bibnamefont{Waddell}}, \bibnamefont{and}
  \bibinfo{author}{\bibfnamefont{D.}~\bibnamefont{Wakeham}},
  \bibinfo{journal}{JHEP} \textbf{\bibinfo{volume}{07}}, \bibinfo{pages}{065}
  (\bibinfo{year}{2019}), \eprint{1810.10601}.

\bibitem[{\citenamefont{Geng et~al.}(2021{\natexlab{c}})\citenamefont{Geng,
  L\"ust, Mishra, and Wakeham}}]{Geng:2021iyq}
\bibinfo{author}{\bibfnamefont{H.}~\bibnamefont{Geng}},
  \bibinfo{author}{\bibfnamefont{S.}~\bibnamefont{L\"ust}},
  \bibinfo{author}{\bibfnamefont{R.~K.} \bibnamefont{Mishra}},
  \bibnamefont{and} \bibinfo{author}{\bibfnamefont{D.}~\bibnamefont{Wakeham}},
  \bibinfo{journal}{JHEP} \textbf{\bibinfo{volume}{08}}, \bibinfo{pages}{003}
  (\bibinfo{year}{2021}{\natexlab{c}}), \eprint{2104.07039}.

\bibitem[{\citenamefont{Kawamoto et~al.}(2022)\citenamefont{Kawamoto, Mori,
  Suzuki, Takayanagi, and Ugajin}}]{Kawamoto2022}
\bibinfo{author}{\bibfnamefont{T.}~\bibnamefont{Kawamoto}},
  \bibinfo{author}{\bibfnamefont{T.}~\bibnamefont{Mori}},
  \bibinfo{author}{\bibfnamefont{Y.-k.} \bibnamefont{Suzuki}},
  \bibinfo{author}{\bibfnamefont{T.}~\bibnamefont{Takayanagi}},
  \bibnamefont{and} \bibinfo{author}{\bibfnamefont{T.}~\bibnamefont{Ugajin}},
  \bibinfo{journal}{JHEP} \textbf{\bibinfo{volume}{05}}, \bibinfo{pages}{060}
  (\bibinfo{year}{2022}), \eprint{2203.03851}.

\bibitem[{\citenamefont{Bianchi et~al.}(2022)\citenamefont{Bianchi, De~Angelis,
  and Meineri}}]{Bianchi2022}
\bibinfo{author}{\bibfnamefont{L.}~\bibnamefont{Bianchi}},
  \bibinfo{author}{\bibfnamefont{S.}~\bibnamefont{De~Angelis}},
  \bibnamefont{and} \bibinfo{author}{\bibfnamefont{M.}~\bibnamefont{Meineri}}
  (\bibinfo{year}{2022}), \eprint{2203.10103}.

\bibitem[{\citenamefont{Suzuki and Takayanagi}(2022)}]{Suzuki2022}
\bibinfo{author}{\bibfnamefont{K.}~\bibnamefont{Suzuki}} \bibnamefont{and}
  \bibinfo{author}{\bibfnamefont{T.}~\bibnamefont{Takayanagi}}
  (\bibinfo{year}{2022}), \eprint{2202.08462}.

\bibitem[{\citenamefont{Kusuki}(2022)}]{Kusuki2022}
\bibinfo{author}{\bibfnamefont{Y.}~\bibnamefont{Kusuki}},
  \bibinfo{journal}{JHEP} \textbf{\bibinfo{volume}{03}}, \bibinfo{pages}{161}
  (\bibinfo{year}{2022}), \eprint{2112.10984}.

\bibitem[{\citenamefont{Anous et~al.}(2022)\citenamefont{Anous, Meineri,
  Pelliconi, and Sonner}}]{Anous2022}
\bibinfo{author}{\bibfnamefont{T.}~\bibnamefont{Anous}},
  \bibinfo{author}{\bibfnamefont{M.}~\bibnamefont{Meineri}},
  \bibinfo{author}{\bibfnamefont{P.}~\bibnamefont{Pelliconi}},
  \bibnamefont{and} \bibinfo{author}{\bibfnamefont{J.}~\bibnamefont{Sonner}}
  (\bibinfo{year}{2022}), \eprint{2202.11718}.

\bibitem[{\citenamefont{Izumi et~al.}(2022)\citenamefont{Izumi, Shiromizu,
  Suzuki, Takayanagi, and Tanahashi}}]{Izumi2022}
\bibinfo{author}{\bibfnamefont{K.}~\bibnamefont{Izumi}},
  \bibinfo{author}{\bibfnamefont{T.}~\bibnamefont{Shiromizu}},
  \bibinfo{author}{\bibfnamefont{K.}~\bibnamefont{Suzuki}},
  \bibinfo{author}{\bibfnamefont{T.}~\bibnamefont{Takayanagi}},
  \bibnamefont{and} \bibinfo{author}{\bibfnamefont{N.}~\bibnamefont{Tanahashi}}
  (\bibinfo{year}{2022}), \eprint{2205.15500}.

\bibitem[{\citenamefont{Chandra et~al.}(2022)\citenamefont{Chandra, Collier,
  Hartman, and Maloney}}]{Chandra2022}
\bibinfo{author}{\bibfnamefont{J.}~\bibnamefont{Chandra}},
  \bibinfo{author}{\bibfnamefont{S.}~\bibnamefont{Collier}},
  \bibinfo{author}{\bibfnamefont{T.}~\bibnamefont{Hartman}}, \bibnamefont{and}
  \bibinfo{author}{\bibfnamefont{A.}~\bibnamefont{Maloney}}
  (\bibinfo{year}{2022}), \eprint{2203.06511}.

\bibitem[{\citenamefont{Belin and de~Boer}(2021)}]{Belin2021a}
\bibinfo{author}{\bibfnamefont{A.}~\bibnamefont{Belin}} \bibnamefont{and}
  \bibinfo{author}{\bibfnamefont{J.}~\bibnamefont{de~Boer}},
  \bibinfo{journal}{Class. Quant. Grav.} \textbf{\bibinfo{volume}{38}},
  \bibinfo{pages}{164001} (\bibinfo{year}{2021}), \eprint{2006.05499}.

\bibitem[{\citenamefont{Cardy}(2004)}]{Cardy2004}
\bibinfo{author}{\bibfnamefont{J.~L.} \bibnamefont{Cardy}}
  (\bibinfo{year}{2004}), \eprint{hep-th/0411189}.

\bibitem[{\citenamefont{Numasawa and Tsiares}(2022)}]{Numasawa2022}
\bibinfo{author}{\bibfnamefont{T.}~\bibnamefont{Numasawa}} \bibnamefont{and}
  \bibinfo{author}{\bibfnamefont{I.}~\bibnamefont{Tsiares}}
  (\bibinfo{year}{2022}), \eprint{2202.01633}.

\bibitem[{\citenamefont{Zamolodchikov and
  Zamolodchikov}(2001)}]{Zamolodchikov2001}
\bibinfo{author}{\bibfnamefont{A.~B.} \bibnamefont{Zamolodchikov}}
  \bibnamefont{and} \bibinfo{author}{\bibfnamefont{A.~B.}
  \bibnamefont{Zamolodchikov}}, pp. \bibinfo{pages}{280--299}
  (\bibinfo{year}{2001}), \eprint{hep-th/0101152}.

\bibitem[{\citenamefont{Hartman et~al.}(2014)\citenamefont{Hartman, Keller, and
  Stoica}}]{Hartman2014}
\bibinfo{author}{\bibfnamefont{T.}~\bibnamefont{Hartman}},
  \bibinfo{author}{\bibfnamefont{C.~A.} \bibnamefont{Keller}},
  \bibnamefont{and} \bibinfo{author}{\bibfnamefont{B.}~\bibnamefont{Stoica}},
  \bibinfo{journal}{JHEP} \textbf{\bibinfo{volume}{09}}, \bibinfo{pages}{118}
  (\bibinfo{year}{2014}), \eprint{1405.5137}.

\bibitem[{\citenamefont{Belin et~al.}(2017)\citenamefont{Belin, Keller, and
  Zadeh}}]{Belin2017}
\bibinfo{author}{\bibfnamefont{A.}~\bibnamefont{Belin}},
  \bibinfo{author}{\bibfnamefont{C.~A.} \bibnamefont{Keller}},
  \bibnamefont{and} \bibinfo{author}{\bibfnamefont{I.~G.} \bibnamefont{Zadeh}},
  \bibinfo{journal}{J. Phys. A} \textbf{\bibinfo{volume}{50}},
  \bibinfo{pages}{435401} (\bibinfo{year}{2017}), \eprint{1704.08250}.

\bibitem[{\citenamefont{Kraus et~al.}(2017)\citenamefont{Kraus,
  Sivaramakrishnan, and Snively}}]{Kraus2017a}
\bibinfo{author}{\bibfnamefont{P.}~\bibnamefont{Kraus}},
  \bibinfo{author}{\bibfnamefont{A.}~\bibnamefont{Sivaramakrishnan}},
  \bibnamefont{and} \bibinfo{author}{\bibfnamefont{R.}~\bibnamefont{Snively}},
  \bibinfo{journal}{JHEP} \textbf{\bibinfo{volume}{08}}, \bibinfo{pages}{084}
  (\bibinfo{year}{2017}), \eprint{1706.00771}.

\bibitem[{\citenamefont{Michel}(2019)}]{Michel2019}
\bibinfo{author}{\bibfnamefont{B.}~\bibnamefont{Michel}}
  (\bibinfo{year}{2019}), \eprint{1908.02873}.

\bibitem[{Note1()}]{Note1}
Note1, \bibinfo{note}{there is an analog of the Hawking-Page transition in
  AdS/BCFT \cite {Fujita2011, Takayanagi2011}.}

\bibitem[{Note2()}]{Note2}
Note2, \bibinfo{note}{we mean cut by inserting the identity operator $\Sigma _p
  \mathinner {|{p}\rangle } \mathinner {\langle {p}|}$.}

\bibitem[{Note3()}]{Note3}
Note3, \bibinfo{note}{we may be able to give the same approximation by
  $\protect \overline {C}_{iip}=0$ and no correlation between $C_{iip}$ and
  $C_{p\protect \mathbb {I}}^a$. One can also realize $C^a_{p \protect \mathbb
  {I}}=0$ by setting $p\protect \neq \protect \bar {p}$. We give further
  comments on this possibility in \cite {Kusuki}.}

\bibitem[{Note4()}]{Note4}
Note4, \bibinfo{note}{there might be a case where we cannot approximate it by
  the vacuum. We will discuss the details in \cite {Kusuki}.}

\bibitem[{\citenamefont{Teschner}(2001)}]{Teschner2001}
\bibinfo{author}{\bibfnamefont{J.}~\bibnamefont{Teschner}},
  \bibinfo{journal}{Class. Quant. Grav.} \textbf{\bibinfo{volume}{18}},
  \bibinfo{pages}{R153} (\bibinfo{year}{2001}), \eprint{hep-th/0104158}.

\bibitem[{\citenamefont{Ponsot and Teschner}(1999)}]{Ponsot1999}
\bibinfo{author}{\bibfnamefont{B.}~\bibnamefont{Ponsot}} \bibnamefont{and}
  \bibinfo{author}{\bibfnamefont{J.}~\bibnamefont{Teschner}}
  (\bibinfo{year}{1999}), \eprint{hep-th/9911110}.

\bibitem[{\citenamefont{Teschner}(2003)}]{Teschner2003}
\bibinfo{author}{\bibfnamefont{J.}~\bibnamefont{Teschner}}, in
  \emph{\bibinfo{booktitle}{{14th International Congress on Mathematical
  Physics}}} (\bibinfo{year}{2003}), \eprint{hep-th/0308031}.

\bibitem[{\citenamefont{Ponsot and Teschner}(2001)}]{Ponsot2001}
\bibinfo{author}{\bibfnamefont{B.}~\bibnamefont{Ponsot}} \bibnamefont{and}
  \bibinfo{author}{\bibfnamefont{J.}~\bibnamefont{Teschner}},
  \bibinfo{journal}{Commun. Math. Phys.} \textbf{\bibinfo{volume}{224}},
  \bibinfo{pages}{613} (\bibinfo{year}{2001}), \eprint{math/0007097}.

\bibitem[{\citenamefont{Kusuki and Miyaji}(2019)}]{Kusuki2019}
\bibinfo{author}{\bibfnamefont{Y.}~\bibnamefont{Kusuki}} \bibnamefont{and}
  \bibinfo{author}{\bibfnamefont{M.}~\bibnamefont{Miyaji}},
  \bibinfo{journal}{JHEP} \textbf{\bibinfo{volume}{08}}, \bibinfo{pages}{063}
  (\bibinfo{year}{2019}), \eprint{1905.02191}.

\bibitem[{\citenamefont{Kusuki}(2019)}]{Kusuki2019a}
\bibinfo{author}{\bibfnamefont{Y.}~\bibnamefont{Kusuki}},
  \bibinfo{journal}{JHEP} \textbf{\bibinfo{volume}{01}}, \bibinfo{pages}{025}
  (\bibinfo{year}{2019}), \eprint{1810.01335}.

\bibitem[{Note5()}]{Note5}
Note5, \bibinfo{note}{this transformation is usually expressed only by an
  integral, but as expressed here, the precise form has an additional
  summation. The summation comes from the poles of the fusion matrix \cite
  {Teschner2001}. One can see the details in, for example, Appendix A of \cite
  {Kusuki2019a}.}

\bibitem[{\citenamefont{Collier et~al.}(2019)\citenamefont{Collier, Gobeil,
  Maxfield, and Perlmutter}}]{Collier2019}
\bibinfo{author}{\bibfnamefont{S.}~\bibnamefont{Collier}},
  \bibinfo{author}{\bibfnamefont{Y.}~\bibnamefont{Gobeil}},
  \bibinfo{author}{\bibfnamefont{H.}~\bibnamefont{Maxfield}}, \bibnamefont{and}
  \bibinfo{author}{\bibfnamefont{E.}~\bibnamefont{Perlmutter}},
  \bibinfo{journal}{JHEP} \textbf{\bibinfo{volume}{05}}, \bibinfo{pages}{212}
  (\bibinfo{year}{2019}), \eprint{1811.05710}.

\bibitem[{\citenamefont{Kusuki and Wei}()}]{Kusuki}
\bibinfo{author}{\bibfnamefont{Y.}~\bibnamefont{Kusuki}} \bibnamefont{and}
  \bibinfo{author}{\bibfnamefont{Z.}~\bibnamefont{Wei}}, \bibinfo{journal}{to
  appear}  (????).

\bibitem[{\citenamefont{Fitzpatrick et~al.}(2014)\citenamefont{Fitzpatrick,
  Kaplan, and Walters}}]{Fitzpatrick2014}
\bibinfo{author}{\bibfnamefont{A.~L.} \bibnamefont{Fitzpatrick}},
  \bibinfo{author}{\bibfnamefont{J.}~\bibnamefont{Kaplan}}, \bibnamefont{and}
  \bibinfo{author}{\bibfnamefont{M.~T.} \bibnamefont{Walters}},
  \bibinfo{journal}{JHEP} \textbf{\bibinfo{volume}{08}}, \bibinfo{pages}{145}
  (\bibinfo{year}{2014}), \eprint{1403.6829}.

\bibitem[{\citenamefont{Miyaji et~al.}(2021)\citenamefont{Miyaji, Takayanagi,
  and Ugajin}}]{Miyaji2021}
\bibinfo{author}{\bibfnamefont{M.}~\bibnamefont{Miyaji}},
  \bibinfo{author}{\bibfnamefont{T.}~\bibnamefont{Takayanagi}},
  \bibnamefont{and} \bibinfo{author}{\bibfnamefont{T.}~\bibnamefont{Ugajin}},
  \bibinfo{journal}{JHEP} \textbf{\bibinfo{volume}{06}}, \bibinfo{pages}{023}
  (\bibinfo{year}{2021}), \eprint{2103.06893}.

\bibitem[{Note6()}]{Note6}
Note6, \bibinfo{note}{the authors in \cite {Miyaji2021} propose that the
  boundary state behaves like a random state in the holographic CFT. But their
  definition of the randomness is quite different from ours.}

\bibitem[{\citenamefont{Collier et~al.}(2020)\citenamefont{Collier, Maloney,
  Maxfield, and Tsiares}}]{Collier2020}
\bibinfo{author}{\bibfnamefont{S.}~\bibnamefont{Collier}},
  \bibinfo{author}{\bibfnamefont{A.}~\bibnamefont{Maloney}},
  \bibinfo{author}{\bibfnamefont{H.}~\bibnamefont{Maxfield}}, \bibnamefont{and}
  \bibinfo{author}{\bibfnamefont{I.}~\bibnamefont{Tsiares}},
  \bibinfo{journal}{JHEP} \textbf{\bibinfo{volume}{07}}, \bibinfo{pages}{074}
  (\bibinfo{year}{2020}), \eprint{1912.00222}.

\bibitem[{Note7()}]{Note7}
Note7, \bibinfo{note}{this SYK model should be modified since it is
  non-trivially coupled to the non-gravitational bath CFT.}

\bibitem[{\citenamefont{Sugiura and Shimizu}(2012)}]{Sugiura2012}
\bibinfo{author}{\bibfnamefont{S.}~\bibnamefont{Sugiura}} \bibnamefont{and}
  \bibinfo{author}{\bibfnamefont{A.}~\bibnamefont{Shimizu}},
  \bibinfo{journal}{Phys. Rev. Lett.} \textbf{\bibinfo{volume}{108}},
  \bibinfo{pages}{240401} (\bibinfo{year}{2012}), \eprint{1112.0740}.

\bibitem[{\citenamefont{Sugiura and Shimizu}(2013)}]{Sugiura2013}
\bibinfo{author}{\bibfnamefont{S.}~\bibnamefont{Sugiura}} \bibnamefont{and}
  \bibinfo{author}{\bibfnamefont{A.}~\bibnamefont{Shimizu}},
  \bibinfo{journal}{Phys. Rev. Lett.} \textbf{\bibinfo{volume}{111}},
  \bibinfo{pages}{010401} (\bibinfo{year}{2013}), \eprint{1302.3138}.

\bibitem[{\citenamefont{Freivogel et~al.}(2021)\citenamefont{Freivogel,
  Nikolakopoulou, and Rotundo}}]{Freivogel2021}
\bibinfo{author}{\bibfnamefont{B.}~\bibnamefont{Freivogel}},
  \bibinfo{author}{\bibfnamefont{D.}~\bibnamefont{Nikolakopoulou}},
  \bibnamefont{and} \bibinfo{author}{\bibfnamefont{A.~F.}
  \bibnamefont{Rotundo}} (\bibinfo{year}{2021}), \eprint{2105.12771}.

\bibitem[{\citenamefont{Goto et~al.}(2021)\citenamefont{Goto, Kusuki, Tamaoka,
  and Ugajin}}]{Goto2021}
\bibinfo{author}{\bibfnamefont{K.}~\bibnamefont{Goto}},
  \bibinfo{author}{\bibfnamefont{Y.}~\bibnamefont{Kusuki}},
  \bibinfo{author}{\bibfnamefont{K.}~\bibnamefont{Tamaoka}}, \bibnamefont{and}
  \bibinfo{author}{\bibfnamefont{T.}~\bibnamefont{Ugajin}},
  \bibinfo{journal}{JHEP} \textbf{\bibinfo{volume}{10}}, \bibinfo{pages}{205}
  (\bibinfo{year}{2021}), \eprint{2108.08308}.

\bibitem[{\citenamefont{Pollack et~al.}(2020)\citenamefont{Pollack, Rozali,
  Sully, and Wakeham}}]{Pollack2020}
\bibinfo{author}{\bibfnamefont{J.}~\bibnamefont{Pollack}},
  \bibinfo{author}{\bibfnamefont{M.}~\bibnamefont{Rozali}},
  \bibinfo{author}{\bibfnamefont{J.}~\bibnamefont{Sully}}, \bibnamefont{and}
  \bibinfo{author}{\bibfnamefont{D.}~\bibnamefont{Wakeham}},
  \bibinfo{journal}{Phys. Rev. Lett.} \textbf{\bibinfo{volume}{125}},
  \bibinfo{pages}{021601} (\bibinfo{year}{2020}), \eprint{2002.02971}.

\bibitem[{\citenamefont{Altland and Sonner}(2021)}]{Altland2021a}
\bibinfo{author}{\bibfnamefont{A.}~\bibnamefont{Altland}} \bibnamefont{and}
  \bibinfo{author}{\bibfnamefont{J.}~\bibnamefont{Sonner}},
  \bibinfo{journal}{SciPost Phys.} \textbf{\bibinfo{volume}{11}},
  \bibinfo{pages}{034} (\bibinfo{year}{2021}), \eprint{2008.02271}.

\bibitem[{\citenamefont{Altland et~al.}(2021)\citenamefont{Altland, Bagrets,
  Nayak, Sonner, and Vielma}}]{Altland2021}
\bibinfo{author}{\bibfnamefont{A.}~\bibnamefont{Altland}},
  \bibinfo{author}{\bibfnamefont{D.}~\bibnamefont{Bagrets}},
  \bibinfo{author}{\bibfnamefont{P.}~\bibnamefont{Nayak}},
  \bibinfo{author}{\bibfnamefont{J.}~\bibnamefont{Sonner}}, \bibnamefont{and}
  \bibinfo{author}{\bibfnamefont{M.}~\bibnamefont{Vielma}},
  \bibinfo{journal}{Phys. Rev. Res.} \textbf{\bibinfo{volume}{3}},
  \bibinfo{pages}{033259} (\bibinfo{year}{2021}), \eprint{2105.12129}.

\bibitem[{\citenamefont{Maloney and Witten}(2020)}]{Maloney2020}
\bibinfo{author}{\bibfnamefont{A.}~\bibnamefont{Maloney}} \bibnamefont{and}
  \bibinfo{author}{\bibfnamefont{E.}~\bibnamefont{Witten}},
  \bibinfo{journal}{JHEP} \textbf{\bibinfo{volume}{10}}, \bibinfo{pages}{187}
  (\bibinfo{year}{2020}), \eprint{2006.04855}.

\bibitem[{\citenamefont{Afkhami-Jeddi et~al.}(2021)\citenamefont{Afkhami-Jeddi,
  Cohn, Hartman, and Tajdini}}]{AfkhamiJeddi2021}
\bibinfo{author}{\bibfnamefont{N.}~\bibnamefont{Afkhami-Jeddi}},
  \bibinfo{author}{\bibfnamefont{H.}~\bibnamefont{Cohn}},
  \bibinfo{author}{\bibfnamefont{T.}~\bibnamefont{Hartman}}, \bibnamefont{and}
  \bibinfo{author}{\bibfnamefont{A.}~\bibnamefont{Tajdini}},
  \bibinfo{journal}{JHEP} \textbf{\bibinfo{volume}{01}}, \bibinfo{pages}{130}
  (\bibinfo{year}{2021}), \eprint{2006.04839}.

\bibitem[{\citenamefont{Janik}(2001)}]{Janik2001}
\bibinfo{author}{\bibfnamefont{R.~A.} \bibnamefont{Janik}},
  \bibinfo{journal}{Nucl. Phys. B} \textbf{\bibinfo{volume}{618}},
  \bibinfo{pages}{675} (\bibinfo{year}{2001}), \eprint{hep-th/0109021}.

\bibitem[{Note8()}]{Note8}
Note8, \bibinfo{note}{it has not been proven that this is the full
  classification of boundaries.}

\bibitem[{\citenamefont{Almheiri
  et~al.}(2020{\natexlab{b}})\citenamefont{Almheiri, Hartman, Maldacena,
  Shaghoulian, and Tajdini}}]{Almheiri2020a}
\bibinfo{author}{\bibfnamefont{A.}~\bibnamefont{Almheiri}},
  \bibinfo{author}{\bibfnamefont{T.}~\bibnamefont{Hartman}},
  \bibinfo{author}{\bibfnamefont{J.}~\bibnamefont{Maldacena}},
  \bibinfo{author}{\bibfnamefont{E.}~\bibnamefont{Shaghoulian}},
  \bibnamefont{and} \bibinfo{author}{\bibfnamefont{A.}~\bibnamefont{Tajdini}},
  \bibinfo{journal}{JHEP} \textbf{\bibinfo{volume}{05}}, \bibinfo{pages}{013}
  (\bibinfo{year}{2020}{\natexlab{b}}), \eprint{1911.12333}.

\end{thebibliography}

\if(
\pagebreak
\widetext
\begin{center}
\textbf{\large Supplemental Materials}
\end{center}
\setcounter{equation}{0}
\setcounter{figure}{0}
\setcounter{table}{0}
\setcounter{page}{1}
\makeatletter
\renewcommand{\theequation}{S\arabic{equation}}
\renewcommand{\thefigure}{S\arabic{figure}}
\renewcommand{\bibnumfmt}[1]{[S#1]}
\renewcommand{\citenumfont}[1]{S#1}

\section{Left-Right Entanglement Entropy for the whole system}

)\fi

\end{document}